\documentclass[11pt]{article}

\usepackage{amssymb,amsmath,graphicx,graphics,color,epstopdf}
\usepackage{authblk}

\usepackage[margin=1in]{geometry}

\usepackage{dblfloatfix}
\usepackage{float}

\usepackage[symbol]{footmisc}
\usepackage[normalem]{ulem}

\newcommand{\de}{\partial}

\newcommand{\eps}{\epsilon}

\newcommand{\etal}{{\em et al.}}

\newcommand{\mn}{\mu \nu}
\newcommand{\rl}{\rho \lambda}
\newcommand{\reff}[1]{(\ref{#1})}

\begin{document}

\title{New Insights into the Nature of Nonlinear Gravitational Waves}
\author{Rosie Hayward\footnote{Corresponding author: rh45@hw.ac.uk} \,and Fabio Biancalana}

\affil{School of Engineering and Physical Sciences, Heriot-Watt University, EH14 4AS Edinburgh, UK}
\maketitle

\begin{abstract}
We study the evolution equations for gravitational waves, which are derived using the full metric to raise and lower indices. This method ensures full consistency between the Ricci tensor and all gauge restrictions and requirements, and allows a meaningful expansion of all tensors up to second order, avoiding several inconsistencies and contradictions observed in previous work. Taking the harmonic gauge to second order in the perturbation theory results in a new nonlinear equation. We show that non-trivial solutions to this equation are necessarily non-plane wave modes with a non-zero trace. These solutions must contain both longitudinal and transverse components, as both are permitted by the gauge restrictions.
\end{abstract}


\textit{Introduction.}--- The weak field approximation in linearised gravity, where a general spacetime metric is decomposed into the flat Minkowski metric and a small perturbation, is well known \cite{Carroll}. From this approximation, the harmonic gauge can be used to show that Einstein's equations reduce to a flat space wave equation for the perturbation; the result is, in standard general relativity (GR), transverse-traceless (TT) plane waves. These gravitational waves present only two modes, which are tensor `cross' and `plus' modes of spin-2 \cite{frgrav1}, coinciding with the spin of the graviton \cite{gravitation}. In this approximation, the second order terms of the Ricci tensor can be calculated to give the energy and momentum of these plane gravitational waves, which correspond to a pseudo stress-energy tensor \cite{Weinberg}. However, in this approximation the first order equation is disconnected from the second order terms, and for gravitational waves propagating in a flat background spacetime it is generally accepted that any influence from nonlinearities can be neglected \cite{Carroll, Weinberg, mendonca2}. 

This is challenged by Aldrovandi  \etal, who argue that linearised gravitational waves cannot represent physical gravitational waves, as gravitational waves carry energy and momentum, their own source \cite{Aldrovandi}. Just as a propagating Yang-Mills field must be nonlinear to carry its own source (colour), a propagating gravitational field must be nonlinear in order to carry energy and momentum. In their paper, they separate the gravitational wave perturbation into different ordered responses, and correspondingly argue that the equation for the second order response should be considered the equation which represents physical gravitational waves.

Nonlinear gravitational waves propagating in a vacuum are also considered in \cite{Canfora}, where the authors contest the assumption that waves arriving at detection facilities such as LIGO and VIRGO can be approximately described by the linearised Einstein field equations. They find a solution with a differing polarisation and spin from the standard result. Alternate theories of gravity, such as $f(R)$ theories, also allow for alternate plane gravitational waves modes to exist, including breathing, longitudinal, and vector modes \cite{frgrav1}. Since LIGO's initial detection of gravitational waves \cite{LIGO}, there have been preliminary tests into their polarisation content, and while pure tensor polarisations are favoured in the data, alternate polarisations have not been ruled out \cite{abbott, pang}. Furthermore, it has been shown that laser interferometer detectors should experience a nonlinear effect from gravitational radiation of equal magnitude to the linear effect \cite{Christodoulou}, and that higher order gravitational wave modes have an important role to play in the detection of this effect \cite{Lasky}. As it stands, the nature of weak-field gravitational wave polarisations are not completely established, and are worth investigating, as is a nonlinear evolution equation for gravitational waves.

We hence believe it should be possible to examine higher order terms in the Ricci tensor, such that one can look for solutions other than transverse-traceless plane waves, and which cannot be reduced to transverse-traceless plane waves by a gauge transformation. In this work, we use the standard perturbation of the gravitational field to calculate the Ricci tensor up to second order in the perturbation. However, our work differs from earlier attempts in a number of ways. 

Firstly, in order to calculate the Ricci tensor to a higher order, one must also calculate the harmonic gauge condition to a higher order, along with all other restrictions. Previously, the second order terms in the Ricci tensor, which are attributed to the pseudo-stress energy tensor of gravitational plane waves, giving their energy and momentum, have been calculated using the harmonic gauge only to first order \cite{Weinberg}. We believe it is logical that all equations relevant to the calculations should be correct to the same order in the gravitational wave perturbation. This means one goes to a higher order when examining the gauge freedoms from coordinate transformations also. 

Secondly, we do not use the background Minkowski metric to raise and lower indices. A gravitational wave is typically treated as a field propagating on a background spacetime, unable to interact with it \cite{donder}. Consequently, the Minkowski metric is considered sufficient when it comes to index manipulation. However, if continuing to higher order in the perturbation, it is necessary to use the full metric to manipulate the indices, otherwise many second-order terms will be lost due to the assumption partial derivatives commute. Instead, we use the full curved-space metric to raise and lower indices, and then cancel away terms of order three or higher. This means partial derivatives in our calculations do not commute. In doing this, we treat our gravitational wave perturbation $h_{\mn}$ as part of the spacetime. One could think of this as similar to including a back-reaction: we allow the gravitational wave to interact with itself. When only taking the Ricci tensor into consideration, there is arguably no difference in results whether one manipulates indices using the flat-space metric, or the full metric. However, the key difference in approach comes in when considering restrictions and definitions: as an example, if considering traceless waves, the condition $h=\eta^{\mn}h_{\mn}=0$ is distinct from  $h=g^{\mn}h_{\mn}=0$. Likewise, the condition $\eta_{\mn}k^{\mu}k^{\nu}=0$ is distinct from the condition $g_{\mn}k^{\mu}k^{\nu}=0$. We argue that if deriving a second order equation, it is the second type of condition which should be used; this is applied to all equations in this work. By manipulating indices with the full metric, we ensure complete consistency between the Ricci tensor and all restrictions equations -- thus avoiding absurd and contradictory results as noticed, for instance, in Refs. \cite{mendonca2}.

Lastly, we consider solutions which are not plane waves, but instead have varying amplitudes. This allows us to find a second order nonlinear equation for gravitational waves. These solutions have the form:
\begin{equation}
h_{\mn}(x)= a_{\mn}(x) e^{i k_{\lambda}x^{\lambda}} + a^*_{\mn}(x) e^{-i k_{\lambda}x^{\lambda}},
\end{equation}
where $k^{\lambda}$ is a wavevector describing the direction of propagation of the gravitational wave, and $a_{\mn}(x)$ is a complex field, analogous the the so-called `envelope' in optics, denoting dependence of the amplitude on the coordinate $x^{\lambda}$. In this manuscript, we assume our gravitational wave travels in the $z$ direction, such that $k^{\lambda}$ has components $k^0=k^3=k$ and $k^1=k^2=0$.

\textit{The Ricci tensor to second order.}--- When the Ricci tensor is usually derived in linearised gravity, the Minkowski metric is used to raise and lower the indices even if terms are kept up to second order \cite{Weinberg, donder}. This approach is standard, where one uses the background spacetime for index manipulation, and ignores the influence of the perturbations. However, when taking nonlinearity into account, we believe this is an oversimplification, and will result in the exclusion of some second order terms: we must consider the effect of the perturbations. If we separate a general metric into a flat spacetime and a perturbation, as in the standard expansion, such that
\begin{equation}
g_{\mn}=\eta_{\mn}+h_{\mn},
\end{equation}
where $\eta_{\mn}$ is the Minkowski metric, then to $\mathcal{O}(h^2)$ the contravariant metric is given by \cite{Choi}
\begin{equation}
g^{\mn}=\tilde{\eta}^{\mn} - h^{\mn} + h^{\mu \lambda}h^{\nu}_{\lambda},
\end{equation}
where $\tilde{\eta}^{\mn}$ is the contravariant Minkowski metric. The above ensures the condition $g^{\mn}g_{\mn}=4$ is upheld. The full metric $g^{\mn}$ should be used when raising indices, and then the unnecessary higher order terms should be discarded. This is why we must distinguish the contravariant Minkowski metric from the covariant: raising the indices on the covariant Minkowski metric using the full metric will not result in the contravariant, due to the fact that the space is curved. As in the standard treatment of gravitational waves, $h_{\mn}$ is assumed to vanish at an infinite distance from the system under study, but it is not assumed small everywhere \cite{Carroll, Weinberg}. The smallness of the perturbation is often managed through use of a small parameter $\eps$, however this parameter has been omitted here in order to improve readability. Its presence is implicit in the equations. It is perhaps worth noting that $h_{\mn}$ is often split into perturbations of differing order,  $g_{\mn}= \eta_{\mn} + \eps h^{(1)}_{\mn} + \eps^2 h^{(2)}_{\mn}  $ representing a first order perturbation and higher order responses \cite{mendonca2, Aldrovandi}. Splitting the final equations up into first and second order is, however, an approximation, and we prefer to consider the perturbation in its entirety by writing it as $h_{\mn}=  \eps h^{(1)}_{\mn} + \eps^2 h^{(2)}_{\mn} $.

We can write the full Ricci tensor in terms of the Christoffel symbols as follows:
\begin{equation} \label{ricchris}
R_{\mu \nu}=\partial_{\rho} \Gamma^{\rho}_{\mu \nu}-\partial_{\nu} \Gamma^{\rho}_{\rho \mu}+ \Gamma^{\rho}_{\rho \lambda} \Gamma^{\lambda}_{\nu \mu}- \Gamma^{\rho}_{\nu \lambda} \Gamma^{\lambda}_{\rho \mu},
\end{equation}
where the Christoffel symbols can be written in terms of the metric,
\begin{equation}
\Gamma^{\rho}_{\mu \nu}=\frac{1}{2}g^{\rho \lambda} ( \partial_{\mu}g_{\lambda \nu} + \partial_{\nu}g_{ \mu \lambda}  -\partial_{\lambda}g_{\mu \nu} ).
\end{equation}
To find our second-order Ricci tensor, we apply the following rule to raise and lower indies: we use the full definition of $g_{\mn}$, and $g^{\mn}$ to second order, and then cancel any terms which are now $\mathcal{O}(h^3)$ and above. This means that the indices of second order terms are effectively raised and lowered by the Minkowski metric, but first order terms are not. Furthermore, this means for first order terms partial derivatives can not be treated as commutative, i.e $\de_{\rho}\de^{\rho}\neq\de^{\rho}\de_{\rho}$. Going beyond using the Minkowski metric to raise and lower indices ensures \emph{all} nonlinear terms are included in the expansion. It is pertinent to ask, if we are not using the background metric, what a raised index means in this work. It means that the contravariant metric is present, to an appropriate order in $h_{\mn}$.

By inserting our definitions of the metric tensor into Equation \ref{ricchris} and simplifying, we find the $\mathcal{O}(h^2)$ Ricci curvature tensor:
\begin{align} \label{riccifinal1}
R_{\mu \nu} =  \frac{1}{2}\bigg[ \de_{\rho}\partial_{\mu}h_{ \nu}^{\rho}- \de_{\rho}\partial^{\rho}h_{\mu \nu} -\de_{\nu}   \partial_{\mu}h_{ \rho}^{\rho} + \de_{\nu}\partial^{\rho}h_{\mu \rho}  \bigg] \\ \nonumber +\frac{1}{2} \bigg[ \de_{\rho} (h_{\lambda \nu}\partial_{\mu}h^{\rho \lambda}+h_{\mu \lambda }\partial_{\nu}h^{\rho \lambda} )  -\de_{\nu}(h_{\lambda \rho}\partial_{\mu}h^{\rho \lambda} + h_{\mu \lambda }\partial_{\rho}h^{\rho \lambda} )\bigg] \\ \nonumber+\frac{1}{4} \bigg[  ( \partial_{\rho}h_{\lambda}^{\rho} + \partial_{\lambda}h_{ \rho }^{\rho}  -\partial^{\rho}h_{\rho \lambda} ) ( \partial_{\mu}h^{\lambda}_{ \nu} + \partial_{\nu}h_{ \mu }^{\lambda}  -\partial^{\lambda} h_{\mu \nu} ) \bigg] \nonumber \\ - \frac{1}{4} \bigg[ ( \partial_{\nu}h_{\lambda}^{\rho} + \partial_{\lambda}h_{ \nu }^{\rho}   -\partial^{\rho} h_{\nu \lambda} )( \partial_{\rho}h_{ \mu}^{\lambda}  + \partial_{\mu}h_{ \rho}^{\lambda}   -\partial^{\lambda} h_{\rho \mu} ) \bigg]  \nonumber
\end{align}

The above can be further simplified through careful manipulation of the higher order terms to give:
\begin{align} \label{riccifinal2}
R_{\mu \nu} =  \frac{1}{2}\bigg[ \de_{\rho}\partial_{\mu}h_{ \nu}^{\rho}- \de_{\rho}\partial^{\rho}h_{\mu \nu} -\de_{\nu}   \partial_{\mu}h_{ \rho}^{\rho} + \de_{\nu}\partial^{\rho}h_{\mu \rho}  \bigg] \\ \nonumber +\frac{1}{2} \bigg[ \de_{\rho} (h_{\lambda \nu}\partial_{\mu}h^{\rho \lambda}+h_{\mu \lambda }\partial_{\nu}h^{\rho \lambda} )  -\de_{\nu}(h_{\lambda \rho}\partial_{\mu}h^{\rho \lambda} + h_{\mu \lambda }\partial_{\rho}h^{\rho \lambda} )\bigg] \\ \nonumber+\frac{1}{4} \bigg[  \partial_{\lambda}h_{ \rho }^{\rho}  ( \partial_{\mu}h^{\lambda}_{ \nu} + \partial_{\nu}h_{ \mu }^{\lambda}  -\partial^{\lambda} h_{\mu \nu} ) \bigg] \nonumber \\ - \frac{1}{4} \bigg[ \de_{\nu} h_{\rl} \de_{\mu}h^{\rl} - 2 \, \de^{\rho} h_{\lambda \nu} \de_{\rho} h^{\lambda}_{\mu} \, + 2 \, \de^{\rho} h_{\lambda \nu} \de^{\lambda} h_{\rho \mu} \bigg]\nonumber
\end{align}

It may seem difficult to define what is meant by `second order in $h$' when indices are raised using perturbation terms. Strictly, we mean second order in the perturbation tensor $h_{\mn}$, such that the `second order' term $h^{\rl}h_{\mn}$ really represents $ \tilde{\eta}^{\rho \alpha} \tilde{\eta}^{\lambda \beta}   h_{\alpha \beta}  h_{\mn}$. Looking at a term from the Ricci tensor:
\begin{equation}
\de_{\rho}\partial_{\mu}h_{ \nu}^{\rho} = \de_{\rho}\partial_{\mu} g^{\rl} h_{ \nu \lambda} \approx  \de_{\rho}\partial_{\mu} \tilde{\eta}^{\rl} h_{ \nu \lambda} -  \de_{\rho}\partial_{\mu} h^{\rl} h_{ \nu \lambda} \approx  \de_{\rho}\partial_{\mu} \tilde{\eta}^{\rl} h_{ \nu \lambda} -  \de_{\rho}\partial_{\mu}  \tilde{\eta}^{\rho \alpha} \tilde{\eta}^{\lambda \beta}   h_{\alpha \beta}  h_{ \nu \lambda}.
\end{equation}

By applying the above formalism to all terms, we can ensure our tensor is truly second order. The Ricci tensor is similar to the result given by Weinberg \cite{Weinberg}, but includes terms which have arisen from our use of the contravariant metric to raise indices. A detailed derivation of Equation \ref{riccifinal2} is given in the Supplementary Material.

\textit{The Harmonic Gauge}--- To remove the ambiguity in the field equations, it is standard to choose a particular coordinate system or gauge. The equations which arise from the gauge restriction will reduce the available degrees of freedom in the gravitational wave tensor $h_{\mn}$ from ten to six. We choose the harmonic coordinate system, which typically appears in gravitational wave research.

The full harmonic gauge condition is
\begin{equation}
g^{\mn}\Gamma_{\mn}^{\rho}=0.
\end{equation}
It is always possible to find a system where the above is true \cite{Weinberg}. The above condition can be written as 
\begin{equation}
\de_{\rho}h^{\rho}_{\nu} -\frac{1}{2} \de_{\nu} h=0 
\end{equation}
to first order in $h_{\mn}$. By moving to the trace reversed frame, it can be shown this is equivalent to the Lorentz gauge, $\de_{\nu}h^{\mn}=0$, which for gravitational plane waves result in the standard restriction that their amplitudes are transverse: there are no components of $h_{\mn}$ which are in the direction of the wavevector $k^{\mu}$. To second order,
\begin{equation}\label{harmg}
 g^{\rho \lambda}\partial^{\nu}h_{\lambda \nu}  - \frac{1}{2}\partial^{\rho}h  - \frac{1}{2}h_{\mn}\partial^{\rho}h^{\mn} =0.
\end{equation}

We can consider the above equation for a traceless wave with varying amplitudes, based on the assumption that it is possible to move to a traceless frame (however, note that trace-reversing will no longer leave us with $\de^{\rho}h_{\rho \nu}=0$, so this is purely an assumption: gravitational waves are not necessarily traceless). For a traceless wave, $h=g^{\mn}h_{\mn}=0$. To second order, it is possible to write $h_{\rl} \de_{\nu}h^{\rl} = \frac{1}{2} \de_{\nu} (h_{\rl} h^{\rl})$, so for a traceless wave with varying amplitudes, both our second order terms will vanish. We are left with the expression:

\begin{equation}\label{harmg}
 g^{\rho \lambda}\partial^{\nu}h_{\lambda \nu} = 0.
\end{equation}

This is somewhat similar to the Lorenz gauge. However, even within the plane wave approximation, this does not force traceless waves to be purely transverse. Purely longitudinal-traceless waves are a valid solution, and will automatically satisfy the tracelessness condition to second order: $g^{\mn}h_{\mn} = \tilde{\eta}^{\mn}h_{\mn} - h^{\mn}h_{\mn} = 0$, as well as the restriction that gravitational waves travel at the speed of light $k^{\mu}k^{\nu}g_{\mn}=0$. For these waves, $a_{00}=a_{33}=-a_{03}$. Interestingly, this solution is very similar to a result given in the work of Aldrovandi \etal\ for the amplitude of a gravitational wave far from the source \cite{Aldrovandi}. The typical form of TT plane waves, where $a_{11}=-a_{22}$ and $a_{12}=a_{21}$ \cite{Weinberg}, will not satisfy the second-order tracelessness condition. In fact, when examining the condition $g^{\mn}h_{\mn} = \tilde{\eta}^{\mn}h_{\mn} - h^{\mn}h_{\mn} = 0$, it seems all purely transverse waves cannot be traceless to second order. In this case, purely transverse plane waves will not automatically satisfy the harmonic gauge. Hence, we conclude that when going beyond linearised gravity, purely longitudinal-traceless (LT) plane waves now represent the simplest possible solution to the harmonic gauge to second order, equation \reff{harmg}. This is unsurprising, as the TT-gauge is known to fail beyond linear order \cite{Harte}.

Outside of the plane wave approximation, waves where the amplitudes $a_{\mn}$ are allowed to depend on the coordinates $x^{\mu}$ will no longer automatically satisfy $\de^{\rho} h_{\rl}=0$. Instead, we will be left with a linear equation for the amplitudes. This is at odds with the understanding that gravitational waves must be inherently nonlinear, and suggests that when we account for varying amplitudes, physical gravitational waves within our framework must not be traceless.  

From here, we can come to the conclusion that for a purely longitudinal gravitational wave mode, the amplitude tensor will have components in the $z$ and $t$ directions which satisfy the condition $a_{03}=-\frac{1}{2} (a_{00} + a_{33})$. This ensures the condition $k^{\mu}k^{\nu}g_{\mn}=0$ is upheld. For a purely transverse gravitational wave mode, the only apparent restriction is that to second order, the mode cannot be traceless. 

From our analysis of the harmonic gauge, it is no longer a requirement that gravitational waves are purely transverse, even for plane waves. If we allow our $a_{\mn}$ to vary, it becomes clear that gravitational waves must be a mixture of transverse and longitudinal modes, with a non-zero trace. Examining restrictions imposed by coordinate invariance is the next important step in determining their nature.

\textit{Coordinate Invariance}--- The small shifts in coordinates, under which Einstein's equations are invariant, must also be taken into account. Considering the effect coordinate invariance will have on $h_{\mn}$ will further lower the available degrees of freedom from six to two. These two degrees of freedom represent the physical gravitational wave. Consider the transformation

\begin{equation}
x^{\mu} \to x'^{\mu}=x^{\mu} + \xi^{\mu}(x).
\end{equation}

The derivative $\de_{\nu}\xi^{\mu}(x)$ is typically taken to be of the same order as $h_{\mn}$. From the definition of the metric in the new coordinate system

\begin{equation}
g'^{\mn}= \frac{\partial x'^{\mu}}{\partial x^{\rho}} \frac{\partial x'^{\nu}}{\partial x^{\lambda}} g^{\rl},
\end{equation}

one can show

\begin{align}\label{htransform}
 h'_{\mn}=  h_{\mn} -g_{\rho \nu} \de_{\mu} \xi^{\rho} -g_{\mu \lambda} \de_{\nu} \xi^{\lambda} + \de_{\nu}\xi^{\rho} \de_{\mu} \xi_{\rho} + \de_{\nu}\xi^{\rho}\de_{\rho}\xi_{\mu} + \de_{\mu}\xi^{\rho}\de_{\rho}\xi_{\nu}.
\end{align}

The standard condition one arrives at in linearised gravity is

\begin{align}\label{atransf}
 a'_{\mn}=  a_{\mn} - k_{\mu} \eps_{\nu} -k_{\nu} \eps_{\mu},
\end{align}

which renders the amplitude components $a_{11}$ and $a_{12}$ unchanged by gauge transformations for a wave propagating in the $z$ direction with wavevector $k^{\mu}=\{k,0,0,k\}^T$, and hence the only components with absolute physical significance \cite{Weinberg}. However, one can show that this is not automatically true in our case.
 
Let us for now assume a general form of our perturbation: $h_{\mn}=a_{\mn}(x) e^{i k_{\lambda}x^{\lambda}}+a^*_{\mn}(x) e^{-i k_{\lambda}x^{\lambda}}$. We take our coordinate shift to have the same nature, so that $\xi^{\mu}(x) = i \zeta^{\mu}(x) e^{i k_{\lambda}x^{\lambda}} - i \zeta^{ \mu*}(x) e^{- i k_{\lambda}x^{\lambda}}$. This is simply an extension of the example in Weinberg, where a constant amplitude shift is chosen for a plane wave \cite{Weinberg}. The change of $h_{\mn}$ under this gauge transformation becomes:
\begin{align}\label{htransform2}
 h'_{\mn}  =  h_{\mn} &-g_{\rho \nu} \de_{\mu} \xi^{\rho} -g_{\mu \lambda} \de_{\nu} \xi^{\lambda} + \de_{\nu}\xi^{\rho} \de_{\mu} \xi_{\rho} + \de_{\nu}\xi^{\rho}\de_{\rho}\xi_{\mu} + \de_{\mu}\xi^{\rho}\de_{\rho}\xi_{\nu} \nonumber \\ 
  = h_{\mn}& - g_{\rho \nu} g_{\mu \lambda}  \bigg[ k^{\lambda} \bar{\xi}^{\rho} + k^{\rho} \bar{\xi}^{\lambda} 
 + i e^{i k_{\lambda}x^{\lambda}} \big( \partial^{\lambda} \zeta^{\rho} + \partial^{\rho} \zeta^{\lambda} \big)  - i e^{-i k_{\lambda}x^{\lambda}} \big( \partial^{\lambda} \zeta^{\rho *} + \partial^{\rho} \zeta^{\lambda *} \big) \bigg] \nonumber \\  &+ \de_{\nu}\xi^{\rho} \de_{\mu} \xi_{\rho} + \de_{\nu}\xi^{\rho}\de_{\rho}\xi_{\mu} + \de_{\mu}\xi^{\rho}\de_{\rho}\xi_{\nu},
\end{align}
where we avoid writing out the full expression for the second order terms, and $\bar{\xi}^{\rho} = - \zeta^{\mu}(x) e^{i k_{\lambda}x^{\lambda}} - \zeta^{ \mu*}(x) e^{- i k_{\lambda}x^{\lambda}}$. If we assume we have plane gravitational waves and a perturbation with a constant amplitude, we can see the components of TT plane waves are no longer automatically unchanged through examination of the first two terms:
\begin{align}
k_{\mu} = g_{\mn} k^{\nu} = g_{\mu 0} k^{0} + g_{\mu 3} k^{3},
\end{align}
and so for a term involving $k_1$ to be zero, one must first assume $h_{10}$ and $h_{13}$ to be zero. This shows that, for plane waves to second order, one can no longer make the argument that $a_{11}$ and $a_{22}$ are the only components with absolute physical significance, as in \cite{Weinberg}.

Let us now go back to \reff{htransform2}, and keep our varying amplitudes. One should be able to see that there will be no components which are automatically unchanged, implying both longitudinal and transverse modes are valid and physical. Hence, we can see that the argument used in Weinberg to assert gravitational waves are purely transverse does not hold to second order, or for waves with varying amplitudes, shown here for the first time. A full derivation of equation \reff{htransform} can be found in the Supplementary Material. From \reff{htransform2}, we conclude there is no apparent preference for either purely longitudinal or purely transverse gravitational wave modes, and therefore true nonlinear gravitational waves contain a mixture of both.

\textit{The Ricci tensor in the harmonic gauge}--- If we insert our harmonic gauge condition given in equation \reff{harmg} into the Ricci tensor \reff{riccifinal2}, it reduces to the following:
\begin{align} \label{ricciharm}
R_{\mu \nu} =  \frac{1}{2}&\bigg[ - \de_{\rho}\partial^{\rho}h_{\mu \nu}   \bigg] \\ \nonumber +\frac{1}{2} &\bigg[ \de_{\rho} h_{\lambda \nu}\partial_{\mu}h^{\rho \lambda}+\de_{\rho}h_{\mu \lambda }\partial_{\nu}h^{\rho \lambda}  - \partial_{\mu} h_{\nu \lambda}\de_{\rho}h^{\rho \lambda}     -\de_{\nu} h_{\mu \lambda }\partial_{\rho}h^{\rho \lambda}   \bigg] \\ \nonumber+\frac{1}{4} &\bigg[  \partial_{\lambda}h_{ \rho }^{\rho}  ( \partial_{\mu}h^{\lambda}_{ \nu} + \partial_{\nu}h_{ \mu }^{\lambda}  -\partial^{\lambda} h_{\mu \nu} ) \bigg] \nonumber \\ - \frac{1}{4}& \bigg[ \de_{\nu} h_{\rl} \de_{\mu}h^{\rl} - 2 \, \de^{\rho} h_{\lambda \nu} \de_{\rho} h^{\lambda}_{\mu} \, + 2 \, \de^{\rho} h_{\lambda \nu} \de^{\lambda} h_{\rho \mu} \bigg].\nonumber
\end{align}

If only the background Minkowski metric is used to raise and lower indices, then the first term reduces to the standard $\Box_{\eta} h_{\mn}$ result. We can separate the first term into linear and nonlinear parts to ensure it appears clearly: 
\begin{align} \label{ricciharm2}
R_{\mu \nu} =  - & \frac{1}{2}\bigg[ \Box_{\eta}h_{ \mn} -    \de_{\rho} ( h^{\rl} \partial_{\lambda}h_{ \mn}   )    \bigg] \\ \nonumber + & \frac{1}{2} \bigg[ \de_{\rho} h_{\lambda \nu}\partial_{\mu}h^{\rho \lambda}+\de_{\rho}h_{\mu \lambda }\partial_{\nu}h^{\rho \lambda}  - \partial_{\mu} h_{\nu \lambda}\de_{\rho}h^{\rho \lambda}     -\de_{\nu} h_{\mu \lambda }\partial_{\rho}h^{\rho \lambda}   \bigg] \\ \nonumber+&\frac{1}{4} \bigg[  \partial_{\lambda}h_{ \rho }^{\rho}  ( \partial_{\mu}h^{\lambda}_{ \nu} + \partial_{\nu}h_{ \mu }^{\lambda}  -\partial^{\lambda} h_{\mu \nu} ) \bigg] \nonumber \\ - &\frac{1}{4} \bigg[ \de_{\nu} h_{\rl} \de_{\mu}h^{\rl} - 2 \, \de^{\rho} h_{\lambda \nu} \de_{\rho} h^{\lambda}_{\mu} \, + 2 \, \de^{\rho} h_{\lambda \nu} \de^{\lambda} h_{\rho \mu} \bigg] \nonumber
\end{align}

In empty space, it is always possible to show $R_{\mn}=0$. Consequently, we find the following nonlinear evolution equation for gravitational waves:
\begin{align} \label{ricciharm3}
\Box_{\eta}h_{ \mn} =   -& \bigg[    \de_{\rho}  h^{\rl} \partial_{\lambda}h_{ \mn}   +    h^{\rl} \de_{\rho} \partial_{\lambda}h_{ \mn}    \bigg] \\ \nonumber + & \bigg[ \de_{\rho} h_{\lambda \nu}\partial_{\mu}h^{\rho \lambda}+\de_{\rho}h_{\mu \lambda }\partial_{\nu}h^{\rho \lambda}  - \partial_{\mu} h_{\nu \lambda}\de_{\rho}h^{\rho \lambda}     -\de_{\nu} h_{\mu \lambda }\partial_{\rho}h^{\rho \lambda}   \bigg] \\ \nonumber+\frac{1}{2} &\bigg[  \partial_{\lambda}h_{ \rho }^{\rho}  ( \partial_{\mu}h^{\lambda}_{ \nu} + \partial_{\nu}h_{ \mu }^{\lambda}  -\partial^{\lambda} h_{\mu \nu} ) \bigg] \nonumber \\ - \frac{1}{2} & \bigg[ \de_{\nu} h_{\rl} \de_{\mu}h^{\rl} - 2 \, \de^{\rho} h_{\lambda \nu} \de_{\rho} h^{\lambda}_{\mu} \, + 2 \, \de^{\rho} h_{\lambda \nu} \de^{\lambda} h_{\rho \mu} \bigg] \nonumber
\end{align}

Expanding the nonlinear terms in \reff{ricciharm3} will reveal factors of $e^{ \pm 2i k_{\lambda}x^{\lambda}}$ which correspond to gravitational second harmonic generation. It is also important to note that nonlinearity is contained in the phase of the waves to second order: 
\begin{equation}
k_{\lambda}x^{\lambda} = g_{\rl}k^{\lambda}x^{\lambda} = \mathbf{k} \cdot \mathbf{x} - \omega t + h_{\rl} k^{\rho}x^{\lambda},
\end{equation}
where $h_{\rl}k^{\rho}x^{\lambda}$ is the nonlinear phase: a signature of entirely nonlinear wave propagation. It is interesting to note the analogies and the differences between Eq. (\ref{ricciharm3}) and the Maxwell wave equation in nonlinear optics. In a nonlinear optical medium the magnitude of the nonlinear refractive index is typically small with respect to the linear one, and can be treated as a small perturbation leading to massive simplifications \cite{agrawal}. Furthermore, the value of the nonlinear refractive index can be tuned by the choice of nonlinear material. In contrast, the medium in which gravitational waves propagate is spacetime, and as is immediately visible in Eq. (\ref{ricciharm3}),  no small parameter can be found in order to simplify the equations further.

It is easy to show that for LT gravitational plane waves, which we have shown are a solution to the harmonic gauge when continuing to second order, both sides of equation \reff{ricciharm3} will vanish. In the case of general solutions to equation \reff{ricciharm3}, finding an exact solution is far more complicated, especially seeing as the harmonic gauge given in equation \reff{harmg} will now contain amplitude derivatives. This challenging task is left for a future publication.

It may seem reasonable to neglect derivatives of the gravitational wave amplitudes for second order terms, as these terms are already typically considered small, and we can assume the amplitudes vary slowly and therefore have small derivatives. This will unfortunately not lead to a nonlinear evolution equation for gravitational waves. As shown in the supplementary material, the resultant second-order terms will vanish for purely longitudinal waves, and for purely transverse waves, the left-hand side of the equation will be disconnected from the right-hand side. This suggests that the derivatives of the higher-order terms need to be taken into account. Neglecting higher order derivatives will work if one separates the gravitational wave perturbation into terms of increasing order, i.e. an initial perturbation and higher order responses, as seen in \cite{Aldrovandi}. However, the same result can be achieved by treating gravitational waves as a combination of waves with transverse and longitudinal modes without needing to use the linearised approximation: under the rules derived in this manuscript this is not forbidden, even for plane waves. Nevertheless, even if an equation which does not completely vanish can be found, any vanishing terms should be replaced with terms involving higher order derivatives: this is vital to the validity of the approximation.

Furthermore, we have shown purely transverse waves cannot be traceless to second order, and due to the fact that trace-reversing will not allow a traceless frame to be found, tracelessness is not seemingly a restriction; this is true whether we allow the $a_{\mn}$ to vary with $x^{\mu}$ or not. In fact, if we allow the $a_{\mn}$ to vary, it seems forcing waves to be traceless would result in the amplitudes obeying a linear equation, something at odds with our fundamental understanding of how gravitational waves must behave.

Our equation is unique from those previously derived up to second order in the gravitational wave perturbation. This is due to the fact we take the harmonic gauge to second order, which will influence the nature of the equation. Our overall results are further impacted by our use of the full metric to manipulate indices, as this will impact definitions such as the trace of the perturbation and the contraction of wavevectors. It will also impact how we define the contravariant perturbation $h^{\mn}$. We believe this approach is correct, as it takes into account that the true spacetime, described by the metric $g_{\mn}=\eta_{\mn}+h_{\mn}$, is a combination of the background spacetime and the gravitational wave: to include higher order terms in the gravitational wave expansion, is to say that we recognise the field is affecting itself, and therefore we should not raise and lower indices with the background spacetime alone. This approach also ensures consistency between the gauge restrictions and the Ricci tensor.  Hence, for a non-plane gravitational wave far from sources, we argue that \reff{ricciharm3} best represents its propagation. This wave must contain both longitudinal and transverse modes, as both are permitted by the gauge restrictions.
\\

\textit{Conclusions}--- We have shown that to second order, a nonlinear evolution equation for gravitational waves can be derived without the assumption that gravitational waves are propagating on a background spacetime, unable to affect it. We have further shown that when considering waves which do not have a constant envelope (i.e. plane waves), tracelessness is too great a restriction, and can not be found. The solutions of these gravitational wave equations must contain both transverse and longitudinal modes. The field equations will yield coupled equations for the components of the gravitational wave modes, which when considered with the harmonic gauge equations, should allow a relationship between the components to be found, and hence a solution. We leave examination of this to future work. It would also be of interest to examine the geodesic deviation equations in our framework, which would provide insight into how gravitational waves would affect the separation of particles. Lastly, it would be of interest to numerically simulate the equations for the gravitational wave modes, and see what insight this provides.
\\

\textit{Acknowledgements}--- The authors would like to acknowledge funding from the EPSRC Centre of Doctoral Training for Condensed Matter Physics (CM-CDT), grant number EP/L015110/1.

\section*{Supplementary Material for the manuscript ``New Insights into the Nature of Nonlinear Gravitational Waves"}

\subsection*{Deriving the Ricci Tensor}

The Ricci tensor can be expressed in terms of Christoffel symbols:

\begin{equation}
R_{\mu \nu}=\partial_{\rho} \Gamma^{\rho}_{\mu \nu}-\partial_{\nu} \Gamma^{\rho}_{\rho \mu}+ \Gamma^{\rho}_{\rho \lambda} \Gamma^{\lambda}_{\nu \mu}- \Gamma^{\rho}_{\nu \lambda} \Gamma^{\lambda}_{\rho \mu},
\end{equation}

which in turn can be expressed in terms of the metric of the spacetime we wish to describe:

\begin{equation}
\Gamma^{\rho}_{\mu \nu}=\frac{1}{2}g^{\rho \lambda} ( \partial_{\mu}g_{\lambda \nu} + \partial_{\nu}g_{ \mu \lambda}  -\partial_{\lambda}g_{\mu \nu} ),
\end{equation}

\begin{equation}
\Gamma^{\rho}_{\rho \mu }=\frac{1}{2}g^{\rho \lambda} ( \partial_{\rho}g_{\lambda \mu} + \partial_{\mu}g_{ \rho \lambda}  -\partial_{\lambda}g_{\rho \mu} ),
\end{equation}

\begin{equation}
\Gamma^{\rho}_{\rho \lambda }=\frac{1}{2}g^{\rho \alpha} ( \partial_{\rho}g_{\alpha \lambda} + \partial_{\lambda}g_{ \rho \alpha}  -\partial_{\alpha}g_{\rho \lambda} ),
\end{equation}

\begin{equation}
\Gamma^{\lambda}_{\mu \nu}=\frac{1}{2}g^{ \lambda \beta} ( \partial_{\mu}g_{\beta \nu} + \partial_{\nu}g_{ \mu \beta}  -\partial_{\beta}g_{\mu \nu} ),
\end{equation}

\begin{equation}
\Gamma^{\rho}_{\nu \lambda }=\frac{1}{2}g^{\rho \alpha} ( \partial_{\nu}g_{\alpha \lambda} + \partial_{\lambda}g_{ \nu \alpha}  -\partial_{\alpha}g_{\nu \lambda} ),
\end{equation}

\begin{equation}
\Gamma^{\lambda}_{\rho \mu }=\frac{1}{2}g^{\lambda \beta} ( \partial_{\rho}g_{\beta \mu} + \partial_{\mu}g_{ \rho \beta}  -\partial_{\beta}g_{\rho \mu} ).
\end{equation}

We can substitute our metric definitions into the Christoffel symbols:

\begin{align}
\Gamma^{\rho}_{\mu \nu}=\frac{1}{2}(\tilde{\eta}^{\rho \lambda}- h^{\rho \lambda}) ( \partial_{\mu}h_{\lambda \nu} + \partial_{\nu}h_{ \mu \lambda}  -\partial_{\lambda}h_{\mu \nu} )
\end{align}

Note that 

\begin{equation}
\Gamma^{\rho}_{\mu \nu}=\Gamma^{\rho}_{\nu \mu}
\end{equation}

and all metrics are symmetric. From here, we can find a definition of $R_{\mu \nu}$ up to order $h^2$:

\begin{align}
R_{\mu \nu} = \partial_{\rho} \bigg[ \frac{1}{2}(\tilde{\eta}^{\rho \lambda}- h^{\rho \lambda}) ( \partial_{\mu}h_{\lambda \nu} + \partial_{\nu}h_{ \mu \lambda}  -\partial_{\lambda}h_{\mu \nu} ) \bigg]\nonumber \\ - \partial_{\nu} \bigg[ \frac{1}{2}(\tilde{\eta}^{\rho \lambda}- h^{\rho \lambda}) ( \partial_{\mu}h_{\lambda \rho} + \partial_{\rho}h_{ \mu \lambda}  -\partial_{\lambda}h_{\mu \rho} ) \bigg]\nonumber \\ +\frac{1}{4} \bigg[ \tilde{\eta}^{\rho \alpha} ( \partial_{\rho}h_{\alpha \lambda} + \partial_{\lambda}h_{ \rho \alpha}  -\partial_{\alpha}h_{\rho \lambda} )\tilde{\eta}^{ \lambda \beta} ( \partial_{\mu}h_{\beta \nu} + \partial_{\nu}h_{ \mu \beta}  -\partial_{\beta}h_{\mu \nu} ) \bigg] \nonumber \\ - \frac{1}{4} \bigg[ \tilde{\eta}^{\rho \alpha} ( \partial_{\nu}h_{\alpha \lambda} + \partial_{\lambda}h_{ \nu \alpha}  -\partial_{\alpha}h_{\nu \lambda} )\tilde{\eta}^{\lambda \beta} ( \partial_{\rho}h_{\beta \mu} + \partial_{\mu}h_{ \rho \beta}  -\partial_{\beta}h_{\rho \mu} ) \bigg].
\end{align}

Raising indices for second order terms:

\begin{align}
R_{\mu \nu} = \partial_{\rho} \bigg[ \frac{1}{2}(\tilde{\eta}^{\rho \lambda}- h^{\rho \lambda}) ( \partial_{\mu}h_{\lambda \nu} + \partial_{\nu}h_{ \mu \lambda}  -\partial_{\lambda}h_{\mu \nu} ) \bigg]&\nonumber \\ - \partial_{\nu} \bigg[ \frac{1}{2}(\tilde{\eta}^{\rho \lambda}- h^{\rho \lambda}) ( \partial_{\mu}h_{\lambda \rho} + \partial_{\rho}h_{ \mu \lambda}  -\partial_{\lambda}h_{\mu \rho} ) \bigg]&\nonumber \\ +\frac{1}{4} \bigg[  ( \partial_{\rho}h_{\lambda}^{\rho} + \partial_{\lambda}h_{ \rho }^{\rho}  -\partial^{\rho}h_{\rho \lambda} ) ( \partial_{\mu}h^{\lambda}_{ \nu} + \partial_{\nu}h_{ \mu }^{\lambda}  -\partial^{\lambda} h_{\mu \nu} ) \bigg]& \nonumber \\ - \frac{1}{4} \bigg[ ( \partial_{\nu}h_{\lambda}^{\rho} + \partial_{\lambda}h_{ \nu }^{\rho}   -\partial^{\rho} h_{\nu \lambda} )( \partial_{\rho}h_{ \mu}^{\lambda}  + \partial_{\mu}h_{ \rho}^{\lambda}   -\partial^{\lambda} h_{\rho \mu} ) \bigg] &.
\end{align}

Raising indices for first order terms: 

\begin{align}
R_{\mu \nu} = \partial_{\rho} \bigg[ \frac{1}{2}( \partial_{\mu}h_{ \nu}^{\rho}+ h_{\lambda \nu}\partial_{\mu}h^{\rho \lambda}+ \partial_{\nu}h_{ \mu }^{\rho}+ h_{\mu \lambda }\partial_{\nu}h^{\rho \lambda}  -\partial^{\rho}h_{\mu \nu} ) \bigg] &
\nonumber \\ - \partial_{\nu} \bigg[ \frac{1}{2}( \partial_{\mu}h_{ \rho}^{\rho}+ h_{\lambda \rho}\partial_{\mu}h^{\rho \lambda}+ \partial_{\rho}h_{ \mu }^{\rho}+ h_{\mu \lambda }\partial_{\rho}h^{\rho \lambda}  -\partial^{\rho}h_{\mu \rho} )\bigg]&
\nonumber \\ +\frac{1}{4} \bigg[  ( \partial_{\rho}h_{\lambda}^{\rho} + \partial_{\lambda}h_{ \rho }^{\rho}  -\partial^{\rho}h_{\rho \lambda} ) ( \partial_{\mu}h^{\lambda}_{ \nu} + \partial_{\nu}h_{ \mu }^{\lambda}  -\partial^{\lambda} h_{\mu \nu} ) \bigg] &\nonumber \\ - \frac{1}{4} \bigg[ ( \partial_{\nu}h_{\lambda}^{\rho} + \partial_{\lambda}h_{ \nu }^{\rho}   -\partial^{\rho} h_{\nu \lambda} )( \partial_{\rho}h_{ \mu}^{\lambda}  + \partial_{\mu}h_{ \rho}^{\lambda}   -\partial^{\lambda} h_{\rho \mu} ) \bigg] &.
\end{align}

Here, we have used the condition $g^{\rl}\partial_{\mu}h_{\lambda \nu} = \partial_{\mu}(g^{\rl}h_{\lambda \nu}) - h_{\lambda \nu} \partial_{\mu}g^{\rl} = \partial_{\mu}(g^{\rl}h_{\lambda \nu}) + h_{\lambda \nu} \partial_{\mu}h^{\rl}$.

Finally, our expression for the Ricci curvature tensor simplifies to:

\begin{align} \label{riccifinal}
R_{\mu \nu} =  \frac{1}{2}\bigg[ \de_{\rho}\partial_{\mu}h_{ \nu}^{\rho}- \de_{\rho}\partial^{\rho}h_{\mu \nu} -\de_{\nu}   \partial_{\mu}h_{ \rho}^{\rho} + \de_{\nu}\partial^{\rho}h_{\mu \rho}  \bigg] &\nonumber \\ +\frac{1}{2} \bigg[ \de_{\rho} (h_{\lambda \nu}\partial_{\mu}h^{\rho \lambda}+h_{\mu \lambda }\partial_{\nu}h^{\rho \lambda} ) -\de_{\nu}(h_{\lambda \rho}\partial_{\mu}h^{\rho \lambda} + h_{\mu \lambda }\partial_{\rho}h^{\rho \lambda} )\bigg] &\nonumber \\ +\frac{1}{4} \bigg[  ( \partial_{\rho}h_{\lambda}^{\rho} + \partial_{\lambda}h_{ \rho }^{\rho}  -\partial^{\rho}h_{\rho \lambda} ) ( \partial_{\mu}h^{\lambda}_{ \nu} + \partial_{\nu}h_{ \mu }^{\lambda}  -\partial^{\lambda} h_{\mu \nu} ) \bigg] &\nonumber \\ - \frac{1}{4} \bigg[ ( \partial_{\nu}h_{\lambda}^{\rho} + \partial_{\lambda}h_{ \nu }^{\rho}   -\partial^{\rho} h_{\nu \lambda} )( \partial_{\rho}h_{ \mu}^{\lambda}  + \partial_{\mu}h_{ \rho}^{\lambda}   -\partial^{\lambda} h_{\rho \mu} ) \bigg] &.
\end{align}

As the indices are raised and lowered by the Minkowski metric for second order terms, we can recognise that some of the terms will cancel:

\begin{align}
R_{\mu \nu} =  \frac{1}{2}\bigg[ \de_{\rho}\partial_{\mu}h_{ \nu}^{\rho}- \de_{\rho}\partial^{\rho}h_{\mu \nu} -\de_{\nu}   \partial_{\mu}h_{ \rho}^{\rho} + \de_{\nu}\partial^{\rho}h_{\mu \rho}  \bigg] &\nonumber \\ +\frac{1}{2} \bigg[ \de_{\rho} (h_{\lambda \nu}\partial_{\mu}h^{\rho \lambda}+h_{\mu \lambda }\partial_{\nu}h^{\rho \lambda} ) -\de_{\nu}(h_{\lambda \rho}\partial_{\mu}h^{\rho \lambda} + h_{\mu \lambda }\partial_{\rho}h^{\rho \lambda} )\bigg] &\nonumber \\ +\frac{1}{4} \bigg[  \partial_{\lambda}h_{ \rho }^{\rho} ( \partial_{\mu}h^{\lambda}_{ \nu} + \partial_{\nu}h_{ \mu }^{\lambda}  -\partial^{\lambda} h_{\mu \nu} ) \bigg] &\nonumber \\ - \frac{1}{4} \bigg[ ( \partial_{\nu}h_{\lambda}^{\rho} + \partial_{\lambda}h_{ \nu }^{\rho}   -\partial^{\rho} h_{\nu \lambda} )( \partial_{\rho}h_{ \mu}^{\lambda}  + \partial_{\mu}h_{ \rho}^{\lambda}   -\partial^{\lambda} h_{\rho \mu} ) \bigg]&,
\end{align}

Expanding the final bracket, by raising, lowering, and relabelling indices, we can show:

\begin{align}
( \partial_{\nu}h_{\lambda}^{\rho} + \partial_{\lambda}h_{ \nu }^{\rho}   -\partial^{\rho} h_{\nu \lambda} )( \partial_{\rho}h_{ \mu}^{\lambda}  + \partial_{\mu}h_{ \rho}^{\lambda}   -\partial^{\lambda} h_{\rho \mu} ) = &\partial_{\nu}h_{\lambda}^{\rho} \partial_{\rho}h_{ \mu}^{\lambda}       +\partial_{\nu}h_{\lambda}^{\rho}  \partial_{\mu}h_{ \rho}^{\lambda}   - \partial_{\nu}h_{\lambda}^{\rho} \partial^{\lambda} h_{\rho \mu} \nonumber \\ +& \partial_{\lambda}h_{ \nu }^{\rho} \partial_{\rho}h_{ \mu}^{\lambda}           + \partial_{\lambda}h_{ \nu }^{\rho}   \partial_{\mu}h_{ \rho}^{\lambda}           - \partial_{\lambda}h_{ \nu }^{\rho} \partial^{\lambda} h_{\rho \mu} \nonumber \\ - & \partial^{\rho} h_{\nu \lambda}\partial_{\rho}h_{ \mu}^{\lambda}  -\partial^{\rho} h_{\nu \lambda} \partial_{\mu}h_{ \rho}^{\lambda} + \partial^{\rho} h_{\nu \lambda}\partial^{\lambda} h_{\rho \mu} \nonumber \\ \nonumber \\
= &\partial_{\nu}h^{\rl} \partial_{\rho}h_{ \mu \lambda}      +\partial_{\nu}h^{\lambda \rho}  \partial_{\mu}h_{ \rl}   - \partial_{\nu}h^{\rl} \partial_{\lambda} h_{\rho \mu} \nonumber \\ + &\partial^{\lambda}h_{ \nu \rho} \partial^{\rho}h_{ \mu \lambda}           + \partial_{\lambda}h_{ \nu \rho}   \partial_{\mu}h^{\rho \lambda}           - \partial_{\lambda}h_{ \nu }^{\rho} \partial^{\lambda} h_{\rho \mu} \nonumber \\ -& \partial^{\rho} h_{\nu \lambda}\partial_{\rho}h_{ \mu}^{\lambda}  -\partial_{\rho} h_{\nu \lambda} \partial_{\mu}h^{ \rho \lambda} + \partial^{\rho} h_{\nu \lambda}\partial^{\lambda} h_{\rho \mu} \nonumber \\ \nonumber \\
= &\partial_{\nu}h^{\lambda \rho}  \partial_{\mu}h_{ \rl}   +2 \partial^{\lambda}h_{ \nu \rho} \partial^{\rho}h_{ \mu \lambda}       - 2\partial_{\lambda}h_{ \nu }^{\rho} \partial^{\lambda} h_{\rho \mu} .
\end{align}

From here, we find our final form of the Ricci tensor:

\begin{align} \label{riccifinal2}
R_{\mu \nu} & =  \frac{1}{2}\bigg[ \de_{\rho}\partial_{\mu}h_{ \nu}^{\rho}- \de_{\rho}\partial^{\rho}h_{\mu \nu} -\de_{\nu}   \partial_{\mu}h_{ \rho}^{\rho} + \de_{\nu}\partial^{\rho}h_{\mu \rho}  \bigg] \\ \nonumber &+\frac{1}{2} \bigg[ \de_{\rho} (h_{\lambda \nu}\partial_{\mu}h^{\rho \lambda}+h_{\mu \lambda }\partial_{\nu}h^{\rho \lambda} )  -\de_{\nu}(h_{\lambda \rho}\partial_{\mu}h^{\rho \lambda} + h_{\mu \lambda }\partial_{\rho}h^{\rho \lambda} )\bigg] \\ \nonumber &+\frac{1}{4} \bigg[  \partial_{\lambda}h_{ \rho }^{\rho}  ( \partial_{\mu}h^{\lambda}_{ \nu} + \partial_{\nu}h_{ \mu }^{\lambda}  -\partial^{\lambda} h_{\mu \nu} ) \bigg] \nonumber \\ & - \frac{1}{4} \bigg[ \de_{\nu} h_{\rl} \de_{\mu}h^{\rl} - 2 \, \de^{\rho} h_{\lambda \nu} \de_{\rho} h^{\lambda}_{\mu} \, + 2 \, \de^{\rho} h_{\lambda \nu} \de^{\lambda} h_{\rho \mu} \bigg]. \nonumber
\end{align}

\newpage

\subsection*{The Harmonic Gauge}

The full harmonic gauge can be expressed as

\begin{equation}
g^{\mn}\Gamma_{\mn}^{\rho}=0,
\end{equation}

which we can expand to find

\begin{equation}
\frac{1}{2}g^{\mn}g^{\rho \lambda} ( \partial_{\mu}g_{\lambda \nu} + \partial_{\nu}g_{ \mu \lambda}  -\partial_{\lambda}g_{\mu \nu} )=0.
\end{equation}

We can raise indices to find

\begin{equation}
  g^{\rho \lambda}\partial^{\nu}g_{\lambda \nu} +   g^{\rho \lambda}\partial^{\mu}g_{ \mu \lambda}  -g^{\mn}\partial^{\rho}g_{\mu \nu} =0,
\end{equation}

which we can simplify to

\begin{equation}
 g^{\rho \lambda}\partial^{\nu}g_{\lambda \nu}  - \frac{1}{2}g^{\mn}\partial^{\rho}g_{\mu \nu} =0.
\end{equation}

The derivative of the Minkowski metric is constant, so we can write,

\begin{equation}
 g^{\rho \lambda}\partial^{\nu}h_{\lambda \nu}  - \frac{1}{2}g^{\mn}\partial^{\rho}h_{\mu \nu} =0.
\end{equation}

We can show that $\partial^{\rho}(g^{\mn}h_{\mu \nu} ) = g^{\mn}\partial^{\rho}h_{\mu \nu} - h_{\mn}\partial^{\rho}h^{\mu \nu}$ , and therefore find

\begin{equation}
 g^{\rho \lambda}\partial^{\nu}h_{\lambda \nu}  - \frac{1}{2}\partial^{\rho}h  - \frac{1}{2}h_{\mn}\partial^{\rho}h^{\mn} =0.
\end{equation}

For now, we shall leave the harmonic gauge in this form.

\subsection*{Ricci Tensor in the Harmonic Gauge}

Earlier, we showed our Ricci tensor had the form

\begin{align} \label{}
R_{\mu \nu} =  \frac{1}{2}\bigg[ \de_{\rho}\partial_{\mu}h_{ \nu}^{\rho}- \de_{\rho}\partial^{\rho}h_{\mu \nu} -\de_{\nu}   \partial_{\mu}h_{ \rho}^{\rho} + \de_{\nu}\partial^{\rho}h_{\mu \rho}  \bigg] &\\ \nonumber +\frac{1}{2} \bigg[ \de_{\rho} (h_{\lambda \nu}\partial_{\mu}h^{\rho \lambda}+h_{\mu \lambda }\partial_{\nu}h^{\rho \lambda} )  -\de_{\nu}(h_{\lambda \rho}\partial_{\mu}h^{\rho \lambda} + h_{\mu \lambda }\partial_{\rho}h^{\rho \lambda} )\bigg]& \\ \nonumber+\frac{1}{4} \bigg[  \partial_{\lambda}h_{ \rho }^{\rho}  ( \partial_{\mu}h^{\lambda}_{ \nu} + \partial_{\nu}h_{ \mu }^{\lambda}  -\partial^{\lambda} h_{\mu \nu} ) \bigg] &\nonumber \\ - \frac{1}{4} \bigg[ \de_{\nu} h_{\rl} \de_{\mu}h^{\rl} - 2 \, \de^{\rho} h_{\lambda \nu} \de_{\rho} h^{\lambda}_{\mu} \, + 2 \, \de^{\rho} h_{\lambda \nu} \de^{\lambda} h_{\rho \mu} \bigg]&. \nonumber
\end{align}

In what follows, we isolate the two terms relevant to the harmonic gauge, and rearrange them to a form which will allow us to make the substitution:

\begin{align} \label{}
R_{\mu \nu} =  \frac{1}{2}\bigg[ \partial_{\mu}( \de_{\rho}h_{ \nu}^{\rho} ) - \de_{\rho}\partial^{\rho}h_{\mu \nu} -\de_{\nu}   \partial_{\mu}h_{ \rho}^{\rho} + \de_{\nu} (\partial^{\rho}h_{\mu \rho} ) \bigg]& \\ \nonumber +\frac{1}{2} \bigg[ \de_{\rho} (h_{\lambda \nu}\partial_{\mu}h^{\rho \lambda}+h_{\mu \lambda }\partial_{\nu}h^{\rho \lambda} )  -\de_{\nu}(h_{\lambda \rho}\partial_{\mu}h^{\rho \lambda} + h_{\mu \lambda }\partial_{\rho}h^{\rho \lambda} )\bigg]& \\ \nonumber+\frac{1}{4} \bigg[  \partial_{\lambda}h_{ \rho }^{\rho}  ( \partial_{\mu}h^{\lambda}_{ \nu} + \partial_{\nu}h_{ \mu }^{\lambda}  -\partial^{\lambda} h_{\mu \nu} ) \bigg] &\nonumber \\ - \frac{1}{4} \bigg[ \de_{\nu} h_{\rl} \de_{\mu}h^{\rl} - 2 \, \de^{\rho} h_{\lambda \nu} \de_{\rho} h^{\lambda}_{\mu} \, + 2 \, \de^{\rho} h_{\lambda \nu} \de^{\lambda} h_{\rho \mu} \bigg]&,\nonumber
\end{align}

\begin{equation}
\partial^{\rho}h_{\mu \rho} = \de_{\lambda}(g^{\rl}h_{\mu \rho}) - h_{\mu \rho} \de_{\lambda} g^{\rl} = \de_{\lambda}h^{\lambda}_{\mu} + h_{\mu \rho} \de_{\lambda} h^{\rl},
\end{equation}

\begin{align} \label{}
R_{\mu \nu} =  \frac{1}{2}\bigg[ \partial_{\mu}( \de_{\rho}h_{ \nu}^{\rho} ) - \de_{\rho}\partial^{\rho}h_{\mu \nu} -\de_{\nu}   \partial_{\mu}h_{ \rho}^{\rho} + \de_{\nu} (\de_{\rho}h^{\rho}_{\mu} + h_{\mu \lambda} \de_{\rho} h^{\rl}) \bigg] & \\ \nonumber +\frac{1}{2} \bigg[ \de_{\rho} (h_{\lambda \nu}\partial_{\mu}h^{\rho \lambda}+h_{\mu \lambda }\partial_{\nu}h^{\rho \lambda} )  -\de_{\nu}(h_{\lambda \rho}\partial_{\mu}h^{\rho \lambda} + h_{\mu \lambda }\partial_{\rho}h^{\rho \lambda} )\bigg] &\\ \nonumber+\frac{1}{4} \bigg[  \partial_{\lambda}h_{ \rho }^{\rho}  ( \partial_{\mu}h^{\lambda}_{ \nu} + \partial_{\nu}h_{ \mu }^{\lambda}  -\partial^{\lambda} h_{\mu \nu} ) \bigg] &\nonumber \\ - \frac{1}{4} \bigg[ \de_{\nu} h_{\rl} \de_{\mu}h^{\rl} - 2 \, \de^{\rho} h_{\lambda \nu} \de_{\rho} h^{\lambda}_{\mu} \, + 2 \, \de^{\rho} h_{\lambda \nu} \de^{\lambda} h_{\rho \mu} \bigg] &,\nonumber
\end{align}

\begin{align}
\de_{\rho}h_{ \nu}^{\rho}= g_{\lambda \nu} \de_{\rho} h^{\rl} +h^{\rl}\de_{\rho} h_{\lambda \nu},
\end{align}

\begin{align} \label{}
R_{\mu \nu} =  \frac{1}{2}\bigg[ \partial_{\mu}( g_{\lambda \nu} \de_{\rho} h^{\rl} +h^{\rl}\de_{\rho} h_{\lambda \nu} ) - \de_{\rho}\partial^{\rho}h_{\mu \nu} -\de_{\nu}   \partial_{\mu}h_{ \rho}^{\rho} \\+ \de_{\nu} (g_{\lambda \mu} \de_{\rho} h^{\rl} +h^{\rl}\de_{\rho} h_{\lambda \mu} + h_{\mu \lambda} \de_{\rho} h^{\rl}) \bigg] &\nonumber \\ \nonumber +\frac{1}{2} \bigg[ \de_{\rho} (h_{\lambda \nu}\partial_{\mu}h^{\rho \lambda}+h_{\mu \lambda }\partial_{\nu}h^{\rho \lambda} )  -\de_{\nu}(h_{\lambda \rho}\partial_{\mu}h^{\rho \lambda} + h_{\mu \lambda }\partial_{\rho}h^{\rho \lambda} )\bigg] &\\ \nonumber+\frac{1}{4} \bigg[  \partial_{\lambda}h_{ \rho }^{\rho}  ( \partial_{\mu}h^{\lambda}_{ \nu} + \partial_{\nu}h_{ \mu }^{\lambda}  -\partial^{\lambda} h_{\mu \nu} ) \bigg] &\nonumber \\ - \frac{1}{4} \bigg[ \de_{\nu} h_{\rl} \de_{\mu}h^{\rl} - 2 \, \de^{\rho} h_{\lambda \nu} \de_{\rho} h^{\lambda}_{\mu} \, + 2 \, \de^{\rho} h_{\lambda \nu} \de^{\lambda} h_{\rho \mu} \bigg] &. \nonumber
\end{align}

Now turning back to the harmonic gauge, we can manipulate the equation, such that it can be substituted into the tensor:

\begin{equation}
 g^{\rho \lambda}\partial^{\nu}h_{\lambda \nu}  - \frac{1}{2}\partial^{\rho}h  - \frac{1}{2}h_{\mn}\partial^{\rho}h^{\mn} =0,
\end{equation}

changing the first term using the product rule:

\begin{equation}
\partial^{\nu}h^{\rho}_{\nu} +h_{\lambda \nu}\de^{\nu}h^{\rho \lambda} - \frac{1}{2}\partial^{\rho}h  - \frac{1}{2}h_{\mn}\partial^{\rho}h^{\mn} =0.
\end{equation}

Using the product rule again:

\begin{equation}
\partial_{\nu}h^{\rho \nu} +h^{\rl}\de^{\nu}h_{\nu \lambda}+h_{\lambda \nu}\de^{\nu}h^{\rho \lambda} - \frac{1}{2}\partial^{\rho}h  - \frac{1}{2}h_{\mn}\partial^{\rho}h^{\mn} =0.
\end{equation}

We can rearrange to find

\begin{equation}
\partial_{\rho}h^{\rho \lambda} =-h^{\lambda \alpha}\de^{\rho}h_{\rho \alpha}-h_{\rho \alpha}\de^{\rho}h^{\alpha \lambda} + \frac{1}{2}\partial^{\lambda}h  + \frac{1}{2}h_{\alpha \rho}\partial^{\lambda}h^{\alpha \rho} 
\end{equation}

which when contacted with $g_{\lambda \nu}$  will give

\begin{equation}
 g_{\lambda \nu} \partial_{\rho}h^{\rho \lambda} =-h_{\nu}^{ \alpha}\de^{\rho}h_{\rho \alpha}-h_{\rho \alpha}\de^{\rho}h^{\alpha}_{\nu} + \frac{1}{2}\partial_{\nu}h  + \frac{1}{2}h_{\alpha \rho}\partial_{\nu}h^{\alpha \rho} .
\end{equation}

We can substitute the above into the first part of the Ricci tensor and simplify:

\begin{align}
\partial_{\mu}( -h_{\nu}^{ \alpha}\de^{\rho}h_{\rho \alpha}-h_{\rho \alpha}\de^{\rho}h^{\alpha}_{\nu} + \frac{1}{2}\partial_{\nu}h  + \frac{1}{2}h_{\alpha \rho}\partial_{\nu}h^{\alpha \rho}  +h^{\rl}\de_{\rho} h_{\lambda \nu} ) - \de_{\rho}\partial^{\rho}h_{\mu \nu} -\de_{\nu}   \partial_{\mu}h_{ \rho}^{\rho} \\+ \de_{\nu} (-h_{\mu}^{ \alpha}\de^{\rho}h_{\rho \alpha}-h_{\rho \alpha}\de^{\rho}h^{\alpha}_{\mu} + \frac{1}{2}\partial_{\mu}h  + \frac{1}{2}h_{\alpha \rho}\partial_{\mu}h^{\alpha \rho} +h^{\rl}\de_{\rho} h_{\lambda \mu} + h_{\mu \lambda} \de_{\rho} h^{\rl}) ,
\end{align}

\begin{align}
\partial_{\mu}( -h_{\nu}^{ \alpha}\de^{\rho}h_{\rho \alpha} + \frac{1}{2}\partial_{\nu}h  + \frac{1}{2}h_{\alpha \rho}\partial_{\nu}h^{\alpha \rho} ) - \de_{\rho}\partial^{\rho}h_{\mu \nu} -\de_{\nu}   \partial_{\mu}h_{ \rho}^{\rho} \\+ \de_{\nu} ( \frac{1}{2}\partial_{\mu}h  + \frac{1}{2}h_{\alpha \rho}\partial_{\mu}h^{\alpha \rho} ) ,
\end{align}

\begin{align}
\partial_{\mu}( -h_{\nu}^{ \alpha}\de^{\rho}h_{\rho \alpha}  + \frac{1}{2}h_{\alpha \rho}\partial_{\nu}h^{\alpha \rho} ) - \de_{\rho}\partial^{\rho}h_{\mu \nu}  + \frac{1}{2}\de_{\nu}(h_{\alpha \rho}\partial_{\mu}h^{\alpha \rho} ).
\end{align}

We can now look again at the full tensor. It should be clear that some of the second order terms will now cancel.

\begin{align} \label{}
R_{\mu \nu} =  \frac{1}{2}\bigg[ \partial_{\mu}( -h_{\nu}^{ \alpha}\de^{\rho}h_{\rho \alpha}  + \frac{1}{2}h_{\alpha \rho}\partial_{\nu}h^{\alpha \rho} ) - \de_{\rho}\partial^{\rho}h_{\mu \nu}  + \frac{1}{2}\de_{\nu}(h_{\alpha \rho}\partial_{\mu}h^{\alpha \rho} )  \bigg] &\\ \nonumber +\frac{1}{2} \bigg[ \de_{\rho} (h_{\lambda \nu}\partial_{\mu}h^{\rho \lambda}+h_{\mu \lambda }\partial_{\nu}h^{\rho \lambda} )  -\de_{\nu}(h_{\lambda \rho}\partial_{\mu}h^{\rho \lambda} + h_{\mu \lambda }\partial_{\rho}h^{\rho \lambda} )\bigg] & \\ \nonumber+\frac{1}{4} \bigg[  \partial_{\lambda}h_{ \rho }^{\rho}  ( \partial_{\mu}h^{\lambda}_{ \nu} + \partial_{\nu}h_{ \mu }^{\lambda}  -\partial^{\lambda} h_{\mu \nu} ) \bigg] &\nonumber \\ - \frac{1}{4} \bigg[ \de_{\nu} h_{\rl} \de_{\mu}h^{\rl} - 2 \, \de^{\rho} h_{\lambda \nu} \de_{\rho} h^{\lambda}_{\mu} \, + 2 \, \de^{\rho} h_{\lambda \nu} \de^{\lambda} h_{\rho \mu} \bigg]&.\nonumber
\end{align}

To second order, $\de_{\mu} (h_{\alpha \rho}\partial_{\nu}h^{\alpha \rho} ) \approx \de_{\nu} (h_{\alpha \rho}\partial_{\mu}h^{\alpha \rho} )$, which allows the terms to cancel:

\begin{align} \label{}
R_{\mu \nu} =  \frac{1}{2}&\bigg[ \partial_{\mu}( -h_{\nu}^{ \alpha}\de^{\rho}h_{\rho \alpha} ) - \de_{\rho}\partial^{\rho}h_{\mu \nu}   \bigg] \\ \nonumber +\frac{1}{2} &\bigg[ \de_{\rho} (h_{\lambda \nu}\partial_{\mu}h^{\rho \lambda}+h_{\mu \lambda }\partial_{\nu}h^{\rho \lambda} )  -\de_{\nu}( h_{\mu \lambda }\partial_{\rho}h^{\rho \lambda} )\bigg] \\ \nonumber+\frac{1}{4}& \bigg[  \partial_{\lambda}h_{ \rho }^{\rho}  ( \partial_{\mu}h^{\lambda}_{ \nu} + \partial_{\nu}h_{ \mu }^{\lambda}  -\partial^{\lambda} h_{\mu \nu} ) \bigg] \nonumber \\ - \frac{1}{4} &\bigg[ \de_{\nu} h_{\rl} \de_{\mu}h^{\rl} - 2 \, \de^{\rho} h_{\lambda \nu} \de_{\rho} h^{\lambda}_{\mu} \, + 2 \, \de^{\rho} h_{\lambda \nu} \de^{\lambda} h_{\rho \mu} \bigg],\nonumber
\end{align}

collecting the other second order terms:

\begin{align} \label{}
R_{\mu \nu} =  \frac{1}{2}&\bigg[ - \de_{\rho}\partial^{\rho}h_{\mu \nu}   \bigg] \\ \nonumber +\frac{1}{2} &\bigg[ \de_{\rho} (h_{\lambda \nu}\partial_{\mu}h^{\rho \lambda}+h_{\mu \lambda }\partial_{\nu}h^{\rho \lambda} ) - \partial_{\mu}( h_{\nu \lambda}\de_{\rho}h^{\rho \lambda} )   -\de_{\nu}( h_{\mu \lambda }\partial_{\rho}h^{\rho \lambda} )\bigg] \\ \nonumber+\frac{1}{4} &\bigg[  \partial_{\lambda}h_{ \rho }^{\rho}  ( \partial_{\mu}h^{\lambda}_{ \nu} + \partial_{\nu}h_{ \mu }^{\lambda}  -\partial^{\lambda} h_{\mu \nu} ) \bigg] \nonumber \\ - \frac{1}{4} &\bigg[ \de_{\nu} h_{\rl} \de_{\mu}h^{\rl} - 2 \, \de^{\rho} h_{\lambda \nu} \de_{\rho} h^{\lambda}_{\mu} \, + 2 \, \de^{\rho} h_{\lambda \nu} \de^{\lambda} h_{\rho \mu} \bigg],\nonumber
\end{align}

and expanding and cancelling, we find

\begin{align} \label{final}
R_{\mu \nu} =  \frac{1}{2} & \bigg[ - \de_{\rho}\partial^{\rho}h_{\mu \nu}   \bigg] \\ \nonumber +\frac{1}{2} & \bigg[ \de_{\rho} h_{\lambda \nu}\partial_{\mu}h^{\rho \lambda}+\de_{\rho}h_{\mu \lambda }\partial_{\nu}h^{\rho \lambda}  - \partial_{\mu} h_{\nu \lambda}\de_{\rho}h^{\rho \lambda}     -\de_{\nu} h_{\mu \lambda }\partial_{\rho}h^{\rho \lambda}   \bigg] \\ \nonumber+\frac{1}{4}& \bigg[  \partial_{\lambda}h_{ \rho }^{\rho}  ( \partial_{\mu}h^{\lambda}_{ \nu} + \partial_{\nu}h_{ \mu }^{\lambda}  -\partial^{\lambda} h_{\mu \nu} ) \bigg] \nonumber \\ - \frac{1}{4}& \bigg[ \de_{\nu} h_{\rl} \de_{\mu}h^{\rl} - 2 \, \de^{\rho} h_{\lambda \nu} \de_{\rho} h^{\lambda}_{\mu} \, + 2 \, \de^{\rho} h_{\lambda \nu} \de^{\lambda} h_{\rho \mu} \bigg].\nonumber
\end{align}

We leave this as the final form of our Ricci tensor in the harmonic gauge.

\subsection*{Gauge Invariance to Second Order}

To move to a new reference frame, the standard formula is

\begin{equation}
g'^{\mn}= \frac{\partial x'^{\mu}}{\partial x^{\rho}} \frac{\partial x'^{\nu}}{\partial x^{\lambda}} g^{\rl}
\end{equation}

where

\begin{equation}
x'^{\mu}=x^{\mu} + \xi (x).
\end{equation}

This means the transformation is

\begin{equation}
 \frac{\partial x'^{\mu}}{\partial x^{\rho}} = \delta^{\mu}_{\rho}+  \frac{\xi^{\mu}(x)}{\partial x^{\rho}}.
\end{equation}

We can infer that the inverse transformation will take the form 

\begin{equation}
 \frac{\partial x^{\nu}} {\partial x'^{\mu}}= \delta^{\nu}_{\mu}-  \frac{\xi^{\nu}(x)}{\partial x^{\mu}} +\mathcal{O}(\xi^2)
\end{equation}

from the fact that like the metric, $\Lambda^{\mu'}_{\nu}\Lambda^{\nu}_{\mu'}=4$, where $\Lambda^{\mu'}_{\nu}= \frac{\partial x'^{\mu}}{\partial x^{\nu}}$. The perturbation $\xi^{\mu}$ is typically assumed to be the same order of magnitude as the perturbation $h_{\mn}$. Hence, when calculating the effect of coordinate shifts on our metric, we must continue to second order, as we have in our other derivations. If we assume the form of the inverse transformation is

\begin{equation}
 \frac{\partial x^{\nu}} {\partial x'^{\mu}}= \delta^{\nu}_{\mu}-  \frac{\xi^{\nu}(x)}{\partial x^{\nu}} + \de_{\mu}\xi^{\rho}(x)\de_{\rho}\xi^{\nu}(x) +\mathcal{O}(\xi^2),
\end{equation}

then the condition,

\begin{equation}
 \frac{\partial x'^{\mu}} {\partial x^{\nu}} \frac{\partial x^{\nu}} {\partial x'^{\mu}}= 4 -   \delta^{\mu}_{\nu} \de_{\mu}\xi^{\nu}(x)  +  \delta_{\mu}^{\nu} \de_{\nu}\xi^{\mu}(x)       +\delta^{\mu}_{\nu} \de_{\mu}\xi^{\rho}(x)\de_{\rho}\xi^{\nu}(x) -  \de_{\mu}\xi^{\nu}(x) \de_{\nu}\xi^{\mu}(x)  + \mathcal{O}(\xi^3) \approx 4,
\end{equation}

is satisfied to second order. We can now derive how the metric will change for a change in coordinates:

\begin{align}
g'_{\mn}= \frac{\partial x^{\rho}}{\partial x'^{\mu}} \frac{\partial x^{\lambda}}{\partial x'^{\nu}} g_{\rl}  = g_{\mn} -g_{\rho \nu} \de_{\mu} \xi^{\rho} -g_{\mu \lambda} \de_{\nu} \xi^{\lambda} + \de_{\nu}\xi^{\rho} \de_{\mu} \xi_{\rho} + \de_{\nu}\xi^{\rho}\de_{\rho}\xi_{\mu} + \de_{\mu}\xi^{\rho}\de_{\rho}\xi_{\nu} 
\end{align}

\begin{align}
g'^{\mn}= g^{\mn} + \de^{\mu} \xi^{\nu} + \de^{\nu} \xi^{\mu} + \de_{\lambda} \xi^{\nu} \de^{\lambda} \xi^{\mu}
\end{align}

If we insert the definition of the covariant and contravariant metric into the above expressions, we can find how our perturbation $h_{\mn}$ will change under coordinate transformations:

\begin{align}
\eta_{\mn} + h'_{\mn}= \eta_{\mn} + h_{\mn} -g_{\rho \nu} \de_{\mu} \xi^{\rho} -g_{\mu \lambda} \de_{\nu} \xi^{\lambda} + \de_{\nu}\xi^{\rho} \de_{\mu} \xi_{\rho} + \de_{\nu}\xi^{\rho}\de_{\rho}\xi_{\mu} + \de_{\mu}\xi^{\rho}\de_{\rho}\xi_{\nu}, 
\end{align}

which simplifies to

\begin{align}
 h'_{\mn}=  h_{\mn} -g_{\rho \nu} \de_{\mu} \xi^{\rho} -g_{\mu \lambda} \de_{\nu} \xi^{\lambda} + \de_{\nu}\xi^{\rho} \de_{\mu} \xi_{\rho} + \de_{\nu}\xi^{\rho}\de_{\rho}\xi_{\mu} + \de_{\mu}\xi^{\rho}\de_{\rho}\xi_{\nu}.
\end{align}

For the contravariant metric:

\begin{align}
\tilde{\eta}^{\mn} - h'^{\mn} + h'^{\mu \lambda} h'^{\nu}_{\lambda}= \tilde{\eta} ^{\mn} - h^{\mn} + h^{\mu \lambda} h'^{\nu}_{\lambda} + \de^{\mu} \xi^{\nu} + \de^{\nu} \xi^{\mu} + \de_{\lambda} \xi^{\nu} \de^{\lambda} \xi^{\mu},
\end{align}

\begin{align}
 h'^{\mn} - h'^{\mu \lambda} h'^{\nu}_{\lambda}=  h^{\mn} -  h^{\mu \lambda} h^{\nu}_{\lambda} - \de^{\mu} \xi^{\nu} - \de^{\nu} \xi^{\mu} - \de_{\lambda} \xi^{\nu} \de^{\lambda} \xi^{\mu}.
\end{align}

We can examine the second order term on the left-hand side to find

\begin{align}
h'^{\mu \lambda} h'^{\nu}_{\lambda} = (h^{\mu \lambda} - \de^{\mu} \xi^{\lambda} - \de^{\lambda} \xi^{\mu})(h^{\nu}_{\lambda} - \de_{\lambda} \xi^{\nu} - \de^{\nu} \xi_{\lambda}) \\ = h^{\mu \lambda} h^{\nu}_{\lambda} - (\de^{\mu} \xi^{\lambda} + \de^{\lambda} \xi^{\mu})h^{\nu}_{\lambda} - (\de_{\lambda} \xi^{\nu} + \de^{\nu} \xi_{\lambda})h^{\mu \lambda},
\end{align}

which gives us an overall expression for the change in $h^{\mn}$:

\begin{align}
 h'^{\mn} =  h^{\mn} - \de^{\mu} \xi^{\nu} - \de^{\nu} \xi^{\mu} - \de_{\lambda} \xi^{\nu} \de^{\lambda} \xi^{\mu} - (\de^{\mu} \xi^{\lambda} + \de^{\lambda} \xi^{\mu})h^{\nu}_{\lambda} -(\de_{\lambda} \xi^{\nu} + \de^{\nu} \xi_{\lambda})h^{\mu \lambda}.
\end{align}

\subsection*{Slowly Varying}

We can make the assumption that the amplitudes of our gravitational wave are slowly varying, and as a result the derivatives of these amplitudes can be treated as small. In this case, for higher order terms containing multiple derivatives, we can approximately describe them as plane waves. For a derivative of the perturbation which is part of a second order term:

\begin{equation}
\de_{\mu}h_{\rho \nu}(x) \approx i k_{\mu} [ a_{\rho \nu}(x) e^{i k_{\lambda}x^{\lambda}} - a^*_{\rho \nu}(x) e^{-i k_{\lambda}x^{\lambda}} ] = i k_{\mu} \bar{h}_{\rho \nu}(x).
\end{equation}

This gives us the following replacements for second order terms:

\begin{equation}
\de_{\lambda} h^{\rl}\de_{\mu}h_{\rho \nu} \approx  -k_{\lambda} k_{\mu} \bar{h}^{\rl}\bar{h}_{\rho \nu},
\end{equation}

\begin{equation}
 h_{\lambda \nu}\de_{\rho}\de_{\mu}h^{\rl} \approx  -k_{\rho} k_{\mu}  h_{\lambda \nu}h^{\rl}.
\end{equation}

We can input these replacements into our final equation,

\begin{align} \label{ricciharm2}
\Box_{\eta}h_{ \mn} =   -& \bigg[    \de_{\rho}  h^{\rl} \partial_{\lambda}h_{ \mn}   +    h^{\rl} \de_{\rho} \partial_{\lambda}h_{ \mn}    \bigg] \\ \nonumber + & \bigg[ \de_{\rho} h_{\lambda \nu}\partial_{\mu}h^{\rho \lambda}+\de_{\rho}h_{\mu \lambda }\partial_{\nu}h^{\rho \lambda}  - \partial_{\mu} h_{\nu \lambda}\de_{\rho}h^{\rho \lambda}     -\de_{\nu} h_{\mu \lambda }\partial_{\rho}h^{\rho \lambda}   \bigg] \\ \nonumber+\frac{1}{2}& \bigg[  \partial_{\lambda}h_{ \rho }^{\rho}  ( \partial_{\mu}h^{\lambda}_{ \nu} + \partial_{\nu}h_{ \mu }^{\lambda}  -\partial^{\lambda} h_{\mu \nu} ) \bigg] \nonumber \\ - \frac{1}{2}& \bigg[ \de_{\nu} h_{\rl} \de_{\mu}h^{\rl} - 2 \, \de^{\rho} h_{\lambda \nu} \de_{\rho} h^{\lambda}_{\mu} \, + 2 \, \de^{\rho} h_{\lambda \nu} \de^{\lambda} h_{\rho \mu} \bigg], \nonumber
\end{align}

to find

\begin{align} \label{ricciharm2}
\Box_{\eta}h_{ \mn} =   -& \bigg[    -k_{\rho}  k_{\lambda} \bar{h}^{\rl}\bar{h}_{ \mn}    -k_{\rho}  k_{\lambda} h^{\rl}h_{ \mn}    \bigg] \\ \nonumber +\frac{1}{2} &\bigg[  -k_{\lambda}k_{\mu} \bar{h}_{ \rho }^{\rho}  h^{\lambda}_{ \nu} -k_\lambda k_{\nu}\bar{h}\bar{h}_{ \mu }^{\lambda}   \bigg] \nonumber \\ - \frac{1}{2} & \bigg[ -k_{\nu}k_{\mu} \bar{h}_{\rl} \bar{h}^{\rl}  \, - 2 \, k^{\rho} k^{\lambda} \bar{h}_{\lambda \nu}  \bar{h}_{\rho \mu} \bigg]. \nonumber
\end{align}

As discussed in the paper, these terms will vanish for longitudinal waves. For transverse waves, the left-hand side cannot match the right-hand side. This suggests the approximation is too extreme, and we must keep higher order terms involving derivatives (as is standard when a term vanishes). It is possible to find a solution to the above equation if one considers gravitational waves to be a mixture of transverse and longitudinal modes, as discussed in the paper. However, for accuracy, one should still account for the vanishing terms, otherwise the approximation is not valid.


\begin{thebibliography}{99}

\bibitem{Carroll} Carroll, S. M. (2019). Spacetime and geometry. Cambridge University Press.

\bibitem{frgrav1} Kausar, H. R., Philippoz, L., \& Jetzer, P. (2016). Gravitational wave polarization modes in f (R) theories. Physical Review D, 93(12), 124071.

\bibitem{gravitation} Misner, C. W., Thorne, K. S., Wheeler, J. A., \& Gravitation, W. H. (1973). Freeman and Co. San Francisco, Calif.

\bibitem{Weinberg} Weinberg, S., \& Dicke, R. H. (1973). Gravitation and cosmology: principles and applications of the general theory of relativity. New York: Wiley.

\bibitem{mendonca2} Servin, M., Marklund, M., Brodin, G., Mendonca, J. T., \& Cardoso, V. (2003). Nonlinear self-interaction of plane gravitational waves. Physical Review D, 67(8), 087501.

\bibitem{donder} Blanchet, L. (2014). Gravitational radiation from post-Newtonian sources and inspiralling compact binaries. Living Reviews in Relativity, 17(1), 2.

\bibitem{LIGO} Abbott, B. P. et al. (LIGO Scientific Collaboration and Virgo Collaboration), Observation of Gravitational Waves from a Binary Black Hole Merger, Phys. Rev. Lett., 116, 061102 (2016). [DOI], [ADS], [arXiv:1602.03837 [gr-qc]].

\bibitem{abbott} Abbott, B. P., Abbott, R., Abbott, T. D., Acernese, F., Ackley, K., Adams, C., ... \& Affeldt, C. (2017). GW170814: a three-detector observation of gravitational waves from a binary black hole coalescence. Physical review letters, 119(14), 141101.

\bibitem{pang} Pang, P. T., Lo, R. K., Wong, I. C., Li, T. G., \& Broeck, C. V. D. (2020). Generic searches for alternative gravitational wave polarizations with networks of interferometric detectors. arXiv preprint arXiv:2003.07375.

\bibitem{Choi}  Choi, S. Y.,  Shim, J. S., and Song, H. S., Phys. Rev. D 51, 2751 (1995). 

\bibitem{agrawal} G. P. Agrawal, {\em Nonlinear Fiber Optics, 5th ed.} (Academic Press, New York, 2013).

\bibitem{Aldrovandi} Aldrovandi, R., Pereira, J. G., da Rocha, R., \& Vu, K. H.  Nonlinear gravitational waves: their form and effects. International Journal of Theoretical Physics, 49(3), 549-563. (2010).

\bibitem{Canfora} Canfora, F., Vilasi, G., \& Vitale, P. (2002). Nonlinear gravitational waves and their polarization. Physics Letters B, 545(3-4), 373-378.

\bibitem{Christodoulou} Christodoulou, D. (1991). Nonlinear nature of gravitation and gravitational-wave experiments. Physical review letters, 67(12), 1486.

\bibitem{Lasky} Lasky, P. D., Thrane, E., Levin, Y., Blackman, J., \& Chen, Y. (2016). Detecting gravitational-wave memory with LIGO: implications of GW150914. Physical review letters, 117(6), 061102.

\bibitem{Harte} Harte, A. I. (2015). Optics in a nonlinear gravitational plane wave. Classical and Quantum Gravity, 32(17), 175017.



\end{thebibliography}
\end{document}